# Long-term evolution of regulatory DNA sequences.
# Part 2: Theory and future challenges.

Elia Mascolo[1], Réka Borbély[1], Noa Ottilie Borst[2], Nicholas H Barton[1], Justin Crocker[2], and Gašper Tkačik[1]

[1]Institute of Science and Technology Austria, Am Campus 1, AT-3400 Klosterneuburg, Austria

[2]Developmental Biology Unit, European Molecular Biology Laboratory, DE-69117 Heidelberg, Germany

**Abstract.** Promoters and enhancers are *cis*-regulatory elements (CREs), DNA sequences that bind transcription factor (TF) proteins to up- or down-regulate target genes. Decades-long efforts yielded TF-DNA interaction models that predict how strongly an individual TF binds arbitrary DNA sequences and how individual binding events on the CRE combine to affect gene expression. These insights can be synthesized into a global, biophysically-realistic, and quantitative genotype-phenotype (GP) map for gene regulation, a "holy grail" for the application of evolutionary theory. A global map provides a rare opportunity to simulate long-term evolution of regulatory sequences and pose several fundamental questions: How long does it take to evolve CREs *de novo*? How many non-trivial regulatory functions exist in sequence space? How connected are they? For which regulatory architecture is CRE evolution most rapid and evolvable? In this article, the second of a two-part series, we review the application of evolutionary concepts – epistasis, robustness, evolvability, tunability, plasticity, and bet-hedging – to the evolution of gene regulatory sequences. We then evaluate the potential for a unifying theory for the evolution of regulatory sequences, and identify key open challenges.

In Part I of this review series [1], we argued that regulatory DNA allows us to reconstruct genotype-phenotype (GP) maps that are both quantitative and global. This opens up an opportunity to advance beyond the trade-off between empirical (yet local and thus incomplete) maps, and global (yet idealized and thus unrealistic) maps. Global and quantitative GP maps for regulatory DNA open the door to studying "long-term evolution", by which we mean evolutionary trajectories that can start from arbitrary DNA sequences (including random non-functional sequences) and proceed indefinitely, with no limit on how many mutations may be accumulated. We contrasted this with "short-term evolution", which only accounts for a few mutations ($\lesssim 10$) around an observed wild-type, for which quantitative phenotype or fitness measurements are currently available.

In Part II of this review series, we discuss how—in addition to biology- and physics-informed modeling and simulations—a general *theory* for the long-term evolution of regulatory sequences and architecture may be within reach. By mapping genotypes into phenotypes, regulatory GP maps act as information-theoretic codes. We review recent developments that bridge information theory and evolutionary theory. These developments begin to address an overarching question: which *regulatory architectures* optimize theoretical objectives? When these architectures themselves can evolve, what do they evolve towards and under what conditions? For an illustration and description of the key notions of *regulatory architecture*, *function*, *grammar* and *code*, we point the reader to the first Figure of Part I of this review series [1].

## I. EPISTASIS, ROBUSTNESS, AND EVOLVABILITY FOR REGULATORY SEQUENCES

In this section, we briefly review the application of the general evolutionary concepts of epistasis, robustness, and evolvability to GP maps for *cis* regulatory sequences. The motivation for this is twofold. First, because such GP maps can be both quantitative and global, these evolutionary concepts can be elevated from qualitative descriptors to computable quantities. Second, as a consequence, these quantities can serve as key ingredients for the development of a more general theory, as we explain in Section II.

**Epistasis** refers to non-additive effects of individual mutations on fitness or on a phenotype of interest. Because the biophysics of TF-DNA interactions necessitates a nonlinear mapping between the TF's binding energy and the resulting gene expression, a non-epistatic, additive expectation would amount to a straw-man null model; a more reasonable baseline should at least capture the non-additive effects of the sigmoid binding nonlinearity. These will induce "global epistasis" [2,3], but also convey substantial predictive power over the effects of combined mutations [4], where the thermodynamic model serves as a null model to understand interactions between mutations [4]. Identifying "specific epistasis" [5–7], i.e. the extra, site-specific epistatic interactions beyond the global ones induced by a single sigmoid, is far more challenging. Specific epistasis could arise because the loci within a single TF binding site contribute to binding energy non-additively [8,9], or because the occupancies of multiple TFs themselves interact non-linearly to control gene expression [10]. Empirical work provides some evidence that specific epistasis can generate rugged landscapes with multiple fitness peaks [11]. It is important, however, to be mindful of methodological issues in these studies: a tight control over statistical and technical errors (e.g., sampling noise, assay variability, batch effects, etc.) is required to ensure that apparent ruggedness reflects genuine epistasis rather than spurious noise-induced artifacts.

When classical toy models of epistasis ignore the sigmoidal transformation from binding energy to gene expression, even an additive system can masquerade as one with hopelessly complex high-order specific epistasis. Explicitly incorporating known physical constraints and the resulting global epistasis substantially reduces the apparent complexity of specific epistasis. This approach has been successfully applied to both TFs and CREs, and is sufficient to explain most of the observed phenotypic variance [5,12,13].

**Robustness** enables the organism to maintain its function in face of perturbations, including mutations, and it is expected to be selectable insofar as it provides a mechanism for preserving fitness. While in idealized asexual models the equilibrium "mutation load" (loss of fitness due to mutation) depends only on deleterious mutation rate [14], in the presence of epistasis or recombination, selection can favor more mutationally robust GP maps that buffer mutational effects [15]. For TF–DNA interactions, the intrinsic binding sigmoid ensures that at sufficiently high TF concentrations, sites with one or a few mismatches can still be occupied, buffering the immediate fitness consequences of point mutations and thereby increasing robustness [96].

Robustness is a broader concept than maintaining function in face of mutations. Gene regulation may also be selected for robustness to environmental and consequent physiological variability. In developmental contexts, the presence of multiple weaker enhancers could ensure proper spatio-temporal gene expression patterns even when the upstream TF concentrations are perturbed [16].

**Evolvability** is the ability of the system to generate phenotypic variation that is heritable and adaptive [17]. Robustness (maintaining existing function) and evolvability (accessing new functions) are often put into contrast, but [18] demonstrated that this need not be the case on a typical TF-DNA GP map, where the neighborhoods of large neutral networks in sequence space are flanked by regulatory sequences binding alternative TFs: this endows the landscapes with local robustness, while larger mutational distances suddenly unlock new phenotypes. These ideas have been applied to landscapes derived from TF-DNA interactions in bacteria [16] and eukaryotes [18–20]. More generally, fitness landscapes where large neutral plateaus connect domains with differing phenotypes can provide high evolvability because cryptic neutral variation enables the population to access various phenotypes without crossing fitness valleys [21]. Alternatively, when gene expression is viewed as a quantitative trait – as opposed to a binary, ON/OFF one – one can ask about the "tunability" of a landscape, i.e., the ability of mutations to fine-tune regulatory function. This property has been reported for both prokaryotic [16] and eukaryotic regulatory logic [19]. Moreover, the evolvability of any specific phenotype can depend strongly on the properties of the GP map. Reconstruction and experimental mapping of an ancient GP map of a TF-DNA regulatory module showed that some DNA specificities are intrinsically far more likely to arise by mutation, and that these biases shaped subsequent evolution of the regulatory module [22]. Both evolvability and robustness are therefore context-dependent properties, affected by the specific TF-CRE interactions involved [16].

Selection for evolvability and robustness might shape regulatory architecture. Evidence from mutagenesis during early metazoan development suggests that promoters are more mutationally robust, with most mutations affecting overall expression level but not their essential function, whereas enhancer mutations readily alter spatiotemporal expression patterns [23,24]. If confirmed, such an "evolutionary division of labor" would represent a candidate architectural strategy that jointly accommodates robustness and evolvability.

Evolvability can be related to, but is distinct from, phenotypic plasticity. While evolvability implies the ability to nimbly adapt to environmental changes via genetic mutations, plasticity refers to non-genetic, yet adaptive (and sometimes rapid and quantitative) responses to environmental change. Plasticity is often used only descriptively, but can be made quantitative in the context of gene regulation, where it naturally maps to the already-introduced concept of a multi-dimensional "regulatory phenotype": the set of "environments" (represented by different intracellular TFs concentrations) paired with their corresponding gene expression responses [13]. Regulation is therefore a form of plasticity where molecular mechanisms adjust gene expression to the environment via a "regulatory function" (see Fig. 1B of Part I of this review series [1]). This has been compared and contrasted with gene expression noise, which generates uncontrolled phenotypic variability and could plausibly be adaptive via "bet hedging" [25], a strategy extensively studied in bacteria [26–28].

## II. TOWARDS AN EVOLUTIONARY THEORY OF GENETIC REGULATORY ARCHITECTURE

The concepts introduced in the previous section can influence the long-term evolution of regulatory architectures, but they are local by construction, being measured around given (often wild-type) sequences. As such, they naturally inform what we have referred to as short-term evolution. A general theory of long-term regulatory evolution must instead capture these notions using quantitative descriptors of the entire GP map. With this perspective in mind, we now turn to what such a general theory should look like, and argue that it should bring together aspects of evolution and population genetics as well as physics and control theory.

On the one hand, physics and control are essential to realistically model *how* regulatory responses are physically generated by molecular components. Regulatory phenotypes are not qualitative traits, but quantitative input-output mappings that describe how gene expression responds to environmental changes. Such mappings are constrained at the systems level by biophysical facts: all concurrently expressed TFs compete for binding to all accessible CREs, and due to limited TF-DNA specificity, any (even non-cognate) TF can bind any CRE, albeit with a sequence-dependent probability, giving rise to some level of unavoidable non-cognate, off-target binding (or

"crosstalk") [29]. Biophysical constraints further dictate that regulatory sequences map into regulatory phenotypes through nonlinear transformations (e.g., due to the sigmoid that maps binding energy onto TF occupancy), so that global epistasis is not an exception but a necessary baseline expectation. Within all these physical and structural constraints, the theory may suggest optimal regulatory strategies under specified objective functions (such as regulatory error minimization, rate-distortion optimality, channel capacity maximization, or description-length minimization). These optima would serve as benchmarks for rationalizing what is possible with existing molecular components and what would be functionally desirable [30,31], regardless of whether such architectures can in fact be produced and maintained by natural selection.

On the other hand, population genetics imposes an entirely orthogonal set of evolutionary constraints: setting limits to how closely and rapidly any such regulatory architectures can actually be approached given available adaptation mechanisms, or maintained in face of mutation and drift. Existing, well-developed theory for evolutionary dynamics [32] allows us to compute or simulate evolutionary outcomes on given GP maps for gene regulation that incorporate said biophysical constraints. At the same time, the quantitative language for expressing these limits and the resulting GP map properties must be provided via the measures of evolvability, robustness, tunability, etc., as discussed in the previous section.

Yet there is a wide epistemological divide between documenting, defining, and quantifying desirable properties of existing GP maps, and between suggesting a first-principles, predictive theory of what gene regulatory GP maps should look like and how they could evolve. There is indeed a venerable history of combining mechanistic constraints and evolutionary arguments to explain gene regulatory architectures; as an early example, we point to Savageau's demand rules formulated in the 1970s, and the follow-up work on the role of activation vs repression [33–35]. Nevertheless, one might argue that, due to the central role of historical contingency in evolution, we cannot hope for any overarching predictive theory for regulatory architectures: we will always be confined to description, perhaps quantitative if we are lucky; yet forever barred from true prediction, even when it is only qualitative.

We strongly disagree. We also point to the similarly optimistic disposition of colleagues pursuing a theory-driven understanding of evolution [36]. Rather than argue the point of principle, we highlight three disjoint but not incompatible lines of inquiry, with the potential to be promoted or combined into a tentative evolutionary theory of gene regulation.

**Direction 1. Evolvable GP maps open new evolutionary trajectories.** The interplay of selection and drift, when acting across genomic loci whose phenotypic effects (the GP map) can themselves evolve, poses an interesting theoretical challenge. If the molecular context is fixed (e.g., TFs binding preferences are immutable, and mutations can only change CRE sequences, moving them closer or further away from the TF motif consensus), the situation is treatable within existing "drift barrier" theory and its extensions [37]: at finite population size, molecular recognition under selection can approach, but typically fails to fully settle at, an optimum. But when mutations can alter features of the GP map itself or how it can be navigated, new evolutionary solutions or pathways can open up. TF promiscuity provides a clear example: it can evolve (through mutations in TF genes altering their binding specificity) and, by altering the shape of the GP map, it can affect the evolutionary timescales and pathways to neofunctionalization [38].

**Direction 2. Information theory for replicated GP maps predicts evolvable gene regulatory architectures.** We recently formalized regulatory sequence evolution as the evolution of co-evolving "replicated genotype–phenotype maps," in which the mapping from regulatory sequences to regulatory phenotypes is itself encoded by evolvable molecular parameters such as transcription factor concentrations, binding preferences, and cooperativities that are shared among multiple CRE loci (hence "replicated") in an organism [10,39]. Under broad conditions, replicated GP maps evolve naturally toward states that maximize the fraction of fit genotypes while minimizing the information that selection must accumulate in the CREs [40,41]. This principle holds, surprisingly, even outside the regime where the CREs evolve under mutations under strong selection ($Ns \to \infty$ limit) [39]. Under realistic biophysical constraints, it enables first-principles predictions of key regulatory parameters, such as TF concentrations, binding site lengths, activation thresholds, and TF cooperativity. The prediction of such features of gene regulatory architecture, which were previously taken as given, highlights the potential of this theoretical development.

This information-theoretic approach also identifies conditions under which entirely new regulatory mechanisms, such as chromatin-based gene silencing, are expected to evolve. This is predicted to happen when the information cost of encoding the novel mechanism in the genome is smaller than the information savings it affords across all regulatory sites. This framework treats bits as a kind of evolutionary currency, allowing the feasibility and eventual fixation of novel mechanisms to be predicted rather than merely described [39].

Lastly, evolutionary simulations of entire CREs show that the architectures predicted by the fitness-information tradeoff [39] may be precisely the most evolvable ones, i.e., those where regulatory function can emerge most rapidly from random sequences [10] (Fig. 1). This link between evolution and information theory frames evolvability as a global and quantifiable feature of the GP map. It also has the potential to explain long-standing puzzles in gene regulation, including the "specificity paradox" [10]: why eukaryotes employ short, cooperatively acting binding sites of diverse strength.

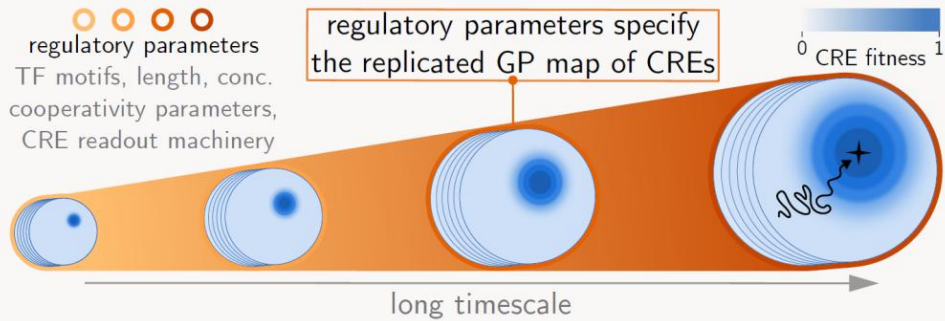

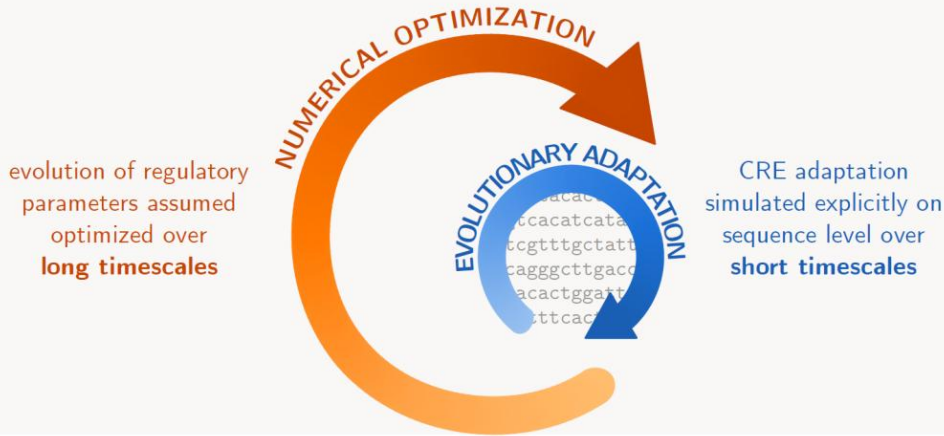

Figure 1. **Co-evolution of CREs and of replicated GP maps for regulatory sequences. (A)** Regulatory parameters (TF concentrations, binding specificities, cooperativity parameters, properties of the non-linear mapping from sequence to expression) are shared among the CREs in the genome, even when individual CREs are selected for different regulatory phenotypes. CREs can thus be seen as evolving on a "replicated GP map" (piled blue discs, with color representing fitness over genotype space). Over very long timescales, evolution favors regulatory parameters (left-to-right progression on the x-axis) that maximize the fraction of fit phenotypes on the replicated GP map for regulatory sequences (the fraction of dark blue increases); at evolutionary equilibrium, this can be understood as a tradeoff between required genetic information to specify individual CREs and mean fitness. Surprisingly, this evolutionary optimization of genetic architecture for replicated maps can occur even when selection acting on individual mutations in CREs is not much stronger than drift. The optimization can predict equilibrium values for regulatory parameters (orange to red) *ab initio* [39]. **(B)** To tractably explore the outcomes of GP map and CRE sequence co-evolution, the "optimize-to-adapt" approach assumes time-scale separation [10]. Shared regulatory parameters change much more slowly than individual CREs evolve. Assuming arguments from (A) hold, we can numerically optimize the regulatory parameters *in silico* (outer orange loop), so that the fraction of fit genotypes in the GP maps is maximized. These equilibrium parameters coincide with those for which evolutionary adaptation of individual CRE sequences (simulated via mutation, selection, and drift;

inner blue loop) is most rapid and, in a sense that can be made mathematically precise, most evolvable.

**Direction 3. Accounting for mutational robustness yields quantitative predictions for regulatory architecture.** Most theoretical and simulational work on the evolution of regulatory sequences and architectures has focused on the weak-mutation regime, where populations are typically fixed for a single genotype and evolve toward a drift-selection balance that can be formalized analogously to the energy-entropy tradeoff of statistical physics. But what happens if the mutation rate is high enough to make populations polymorphic at CRE loci (a regime most likely to be relevant for prokaryotes)? The link between evolution and information theory naturally extends to this high-mutation regime: mutational flux acts as an additional entropic force, and selection must now accumulate a predictable amount of extra information to select genotypes that are not just functional but also more mutationally robust [39]. The action of selection for robustness thus becomes a quantifiable global metric, measurable in bits. Much work remains to flesh out such theoretical extensions and their implications.

An early indication that such extensions are worthwhile is provided by the elegant work by Stewart and colleagues, who argue why TF motifs are around 10bp long [42]. This work incorporates many attributes that we believe would be important for a more general theory: it takes into account biophysical constraints and nonlinearities, it is formulated at the systems-level to model crosstalk, and it considers not only the evolution of CREs but also of the GP maps themselves (e.g., motif lengths can evolve). The $\sim$10bp TF length sweet spot is identified by the theory as a balance between mutational robustness and the TF specificity required to implement reliable gene regulation. While this argument does not explain differences between prokaryotes and eukaryotes and the "specificity paradox," it showcases that including mutational robustness within a broader theoretical framework may be essential.

### III. FUTURE CHALLENGES

We conclude this review series by highlighting several outstanding theoretical and simulational challenges that now appear increasingly tractable. For brevity, we do not dwell on the (perfectly valid!) efforts to scale up all existing directions, from collecting more data on regulatory GP maps and analyzing them with better dimensionality-reduction methods, to refining mechanistic models with ever more detailed knowledge. Rather, we highlight higher-level arguments concerning the evolution of the regulatory system as a whole.

**Challenge 1. What are the most informative sequence ensembles to pursue, theoretically and experimentally?** In Part I of this review series, we examined how classical mutagenesis studies on wild-type regulatory sequences can be complemented by a more recent focus on fully random libraries, which are closely linked to evolutionary theory [13,41]. Importantly, they can offer quantitative estimates of robustness (Fig. 2A) and evolvability (Fig. 2B-C) as described in this review, i.e., as a feature of global rather than local GP maps. But are there library designs that interpolate between these two extremes? Are there clever ways to use phylogeny or theory to guide genotype construction for massively parallel assays? The iterative and directed enrichment of initially random libraries has already been used for synthetic promoter design [43], in SELEX-inspired methods to derive TF binding preferences [44], or for optimizing gene regulatory phenotypes [26]. Such directed (supervised) iterative approaches could be extended into the evolutionary domain, where more exploratory (unsupervised) yet still model-informed methods could allow experiments to efficiently navigate the vast sequence space to better extract global, quantitative GP maps.

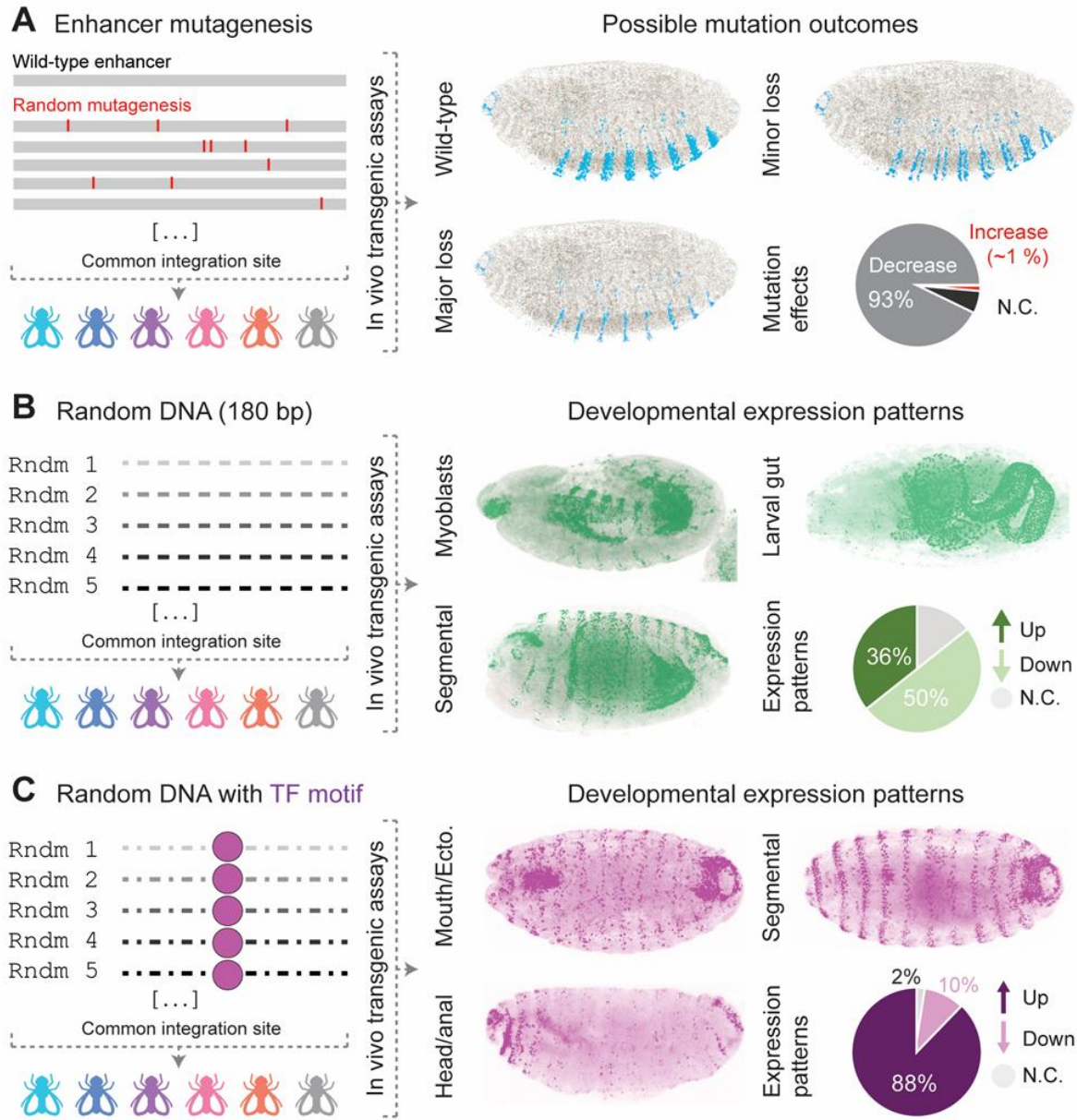

Figure 2. **Regulatory potential revealed by enhancer mutagenesis and random DNA in *Drosophila melanogaster*.** Data from [45]. **(A)** Low mutational robustness of developmental enhancers. Random mutagenesis of native enhancers cloned upstream of a reporter gene, followed by site-specific integration and reporter assays in stage 15 embryos, shows that most mutations reduce enhancer output. For the shavenbaby *E3N* enhancer, most variants drive decreased expression, highlighting the sensitivity of developmental enhancers to sequence perturbations. **(B-C)** Expression is scored relative to a minimal reporter as increased (up), decreased (down), or unchanged (N.C.). **(B)** Latent regulatory activity in random DNA. Fully random 180 bp sequences cloned upstream of a reporter gene, when integrated at a common genomic site, frequently drive reproducible and spatially patterned reporter gene expression in late embryos, demonstrating that regulatory activity is widespread in sequence space even in the absence of evolved enhancers. **(C)** TF motifs enhance the evolvability of new regulatory function from random sequences. Random DNA libraries biased toward binding sites for the pioneer factor Zelda show a marked increase in both the frequency and magnitude of developmental expression, indicating that incorporation of key TF motifs strongly biases random sequence toward enhancer-like behavior.

**Challenge 2. Isolating the impact of specific regulatory mechanisms on long-term sequence evolution.** Are the molecular mechanisms (and corresponding GP map features) with predictive power near the extant wild-type sequences the same mechanisms (and GP map features) that govern long-term regulatory sequence evolution? For instance, [8] found that promiscuous RNAP binding with variable spacer length on bacterial promoters is essential for the predicted long-term evolutionary dynamics, as it accelerated predicted rates of *de novo* regulatory function emergence by orders of magnitude. Yet, when studying extant *Escherichia coli* promoters, this mechanistic feature seems much less important – omitting it causes a significant but quite small drop in predictive power. In other words, features that are essential for modeling regulatory sequences sampled around existing wild-type CREs may differ from those required to recapitulate long-term adaptive dynamics. Seen in a positive light, this consideration puts forward the (simulated) evolutionary dynamics as a possibly new "arbiter" for distinguishing essential from incremental mechanisms (and GP map features).

More broadly, this Challenge forces us to think about how to link, step-by-step, two different levels of description. On the one hand, we have molecular mechanisms and the statistical features that they induce on GP maps inferred from massively parallel assays. On the other hand, we have well-established but phenomenological concepts and quantities of evolutionary biology, such as fitness landscapes, genetic information, distributions of fitness effects, etc. How are these levels of description related? Concretely, if we had access to the detailed, global and quantitative GP map, would we know how to "coarse-grain" it to a much-simpler-to-treat adaptive process that unfolds solely in the phenotypic or fitness space? If possible and done correctly, such coarse-graining would hide the complexity of the underlying high-dimensional GP map and enable efficient simulation, while still quantitatively capturing adaptation timescales and dynamics. In other domains of population genetics, such approaches are already well-established: we point to quantitative genetics and the infinitesimal model as excellent examples [13,46].

**Challenge 3**. **Can information theory quantify regulatory code optimality?** Shannon's information theory has piqued the interest of evolutionary biologists very early [47], but has remained on the sidelines of the mainstream, in contrast to its role in other branches of life science [48,49]. Recently established theoretical links suggest that, absent mechanistic constraints and under strong selection, regulatory architecture would evolve towards CREs that implement a Shannon-optimal code for desired phenotypes [39,41]. While more work is needed to grasp the impact and breadth of this connection, the simple fact that it links two fundamental theories, i.e., evolution by natural selection and information theory, suggests that notions thought to be merely descriptive for genomes (e.g., sequence information, redundancy, compression, coding efficiency, etc.) may provide deeper quantitative insights, at least when applied to regulatory sequences.

**Challenge 4**. **Evolution of entire gene regulatory circuits for emergent, organismal-level phenotypes.** Gene regulation is a strongly interacting, feedback-rich, networked system, encoded by all CREs. Consequently, the function and evolutionary fate of an individual CRE may not be understandable when analyzed as an isolated input/output "device". While it is not unreasonable to assume, as a first approximation, that high fitness should correlate with precise and timely response of such devices to environmental (i.e., TF concentration) changes, it is ultimately the output of the entire gene regulatory network that gives rise to biological function and that is selected for. In this review series, we focused on efforts to bridge the gap between CRE sequence and regulatory (gene expression) phenotype; Challenge 4 argues for further linking gene expression phenotypes (plural!) of multiple interacting CREs to collective, higher-order phenotypic functions. Examples here might include: the emergence of tissue or organismal shape and geometry, controlled by multiple genes and mediated by tissue mechanics; or cell-cell communication modulating gene regulatory network interactions across multiple cells that give rise to a reproducible body plan during early organismal development; or the gene regulatory networks in bacteria controlling enzyme levels to keep cellular metabolism running efficiently across different environments.

Impressive work has been carried out, especially in the evo-devo and biophysics communities, on the evolution of circuits giving rise to observed tissue- or organism-scale organization [50–55]. Yet most of this work abstracted away high-dimensional regulatory sequences, focused on mutational moves taking place directly at the level of gene regulatory network parameters, and (often but not always) replaced evolutionary with numerical optimization. While this drastically reduces the dimensionality of the search space and makes such complex problems tractable simulation-wise [56], it also closes the door to the rigorous and quantitative application of evolvability, robustness, and genetic information concepts, as well as to realistic estimates of adaptation speed, which are only truly calibrated with reference to the actual genotypic space and mutational moves taking place therein. A grand vision for the future is a quantitative framework that connects regulatory sequences, intermediate phenotypes (e.g., gene expression levels, regulatory network parameters, etc.), emerging tissue- or organism-level phenotypes (e.g., tissue patterning and morphology) and organismal fitness. This vision is ambitious because key links remain poorly understood, from long-range and 3D regulatory interactions to enhancer-promoter specificity and the propagation of expression changes through developmentally integrated, pleiotropic, and polygenic trait architectures. A somewhat less grand but more tractable goal, still with a huge impact across multiple fields, would be to establish such a connection for at least one, or perhaps a few, concrete biological systems.

**Acknowledgments**

We thank Calin Guet and Santiago Herrera-Álvarez for essential contributions to this manuscript. E.M. acknowledges support from the APART-USA fellowship, jointly funded by the Austrian Academy of Sciences (ÖAW) and the Institute of Science and Technology Austria (ISTA). N.B. acknowledges funding from the ERC Advanced Grant 101055327 "HaplotypeStructure".

This study was supported by the European Molecular Biology Laboratory (N.O.B., J.C.); the European Molecular Biology Laboratory Interdisciplinary Postdoc Programme (EIPOD) under the Marie Skłodowska-Curie Actions cofund (S.H.A.).